\documentclass[12pt]{article}
\usepackage{latexsym}
\usepackage{cite}

\newtheorem{theorem}{Theorem}

\begin{document}

\begin{center}
{\Large \bf  Lift of noninvariant solutions of heavenly equations
from three to four dimensions and new ultra-hyperbolic metrics}\\[4mm]
{\large \bf A A Malykh$^1$, Y Nutku$^2$ and
M B Sheftel$^{1,2}$}\\[3mm]
$^1$ Department of Higher Mathematics, North Western State
Technical University, Millionnaya St. 5, 191186, St. Petersburg,
Russia
\\ $^2$ Feza G\"{u}rsey Institute, PO Box 6, Cengelkoy,
81220 Istanbul, Turkey \vspace{1mm}
\\ E-mail: specarm@mail.wplus.net, nutku@gursey.gov.tr,
mikhail.sheftel@boun.edu.tr
\end{center}

\begin{abstract}
 We demonstrate that partner symmetries provide a lift of noninvariant solutions
of three-dimensional Boyer-Finley equation to noninvariant
solutions of four-dimensional hyperbolic complex Monge-Amp\`ere
equation. The lift is applied to noninvariant solutions of the
Boyer-Finley equation, obtained earlier by the method of group
foliation, to yield noninvariant solutions of the hyperbolic
complex Monge-Amp\`ere equation. Using these solutions we
construct new Ricci-flat ultra-hyperbolic metrics with non-zero
curvature tensor that have no Killing vectors.
\end{abstract}
AMS classification scheme
numbers: 35Q75, 83C15

\section{Introduction}
\setcounter{equation}{0}

In his paper \cite{pleb} Pleba\~nski introduced his first and
second heavenly equations for a single potential governing
Ricci-flat metrics on $4$-dimensional complex manifolds. Solutions
of these equations determine (anti-)self-dual heavenly metrics
which satisfy the complex vacuum Einstein equations. There are two
real cross sections of the complex metrics governed by the first
heavenly equation, namely K\"ahler metrics with Euclidean or
ultra-hyperbolic signature. The first heavenly equation in these
cases coincides with the elliptic and hyperbolic complex
Monge-Amp\`ere equation ($CMA$) respectively that have
applications to important problems in physics and geometry. In
particular, some solutions $u = u(z_1,\bar z_1,z_2,\bar z_2)$ to
the elliptic $CMA$
\begin{equation}
u_{1\bar 1} u_{2\bar 2} - u_{1\bar 2} u_{2\bar 1} = 1
\label{cma}
\end{equation}
can be interpreted as gravitational instantons. (From now on
subscripts denote partial derivatives with respect to
corresponding variables and bars mean complex conjugates.) The
most important gravitational instanton is the Kummer surface $K3$
\cite{ahs}. The explicit construction of $K3$ metric is still an
unsolved challenging problem. One of the basic difficulties is
that the metric should have no Killing vectors and hence the
corresponding solution of $CMA$ should have no symmetries, i.e. be
a noninvariant solution. That means that the traditional method of
Lie symmetry reduction cannot be applied for finding solutions of
the heavenly equations and therefore there is a problem of finding
their noninvariant solutions. We have recently developed the
method of partner symmetries appropriate for this problem and
obtained certain classes of noninvariant solutions to the elliptic
and hyperbolic $CMA$ and second heavenly equation together with
corresponding heavenly metrics with no Killing vectors
\cite{mns,mnsgr,mnsh}.

In this paper we obtain new noninvariant solutions of the
four-dimensional hyperbolic complex Monge-Amp\`ere equation
($HCMA$)
\begin{equation}
u_{1\bar 1} u_{2\bar 2} - u_{1\bar 2} u_{2\bar 1} = -1
\label{hcma}
\end{equation}
by lifting noninvariant solutions of the three-dimensional
Boyer-Finley equation \cite{bf} to a four-dimensional solution
manifold of $HCMA$. Noninvariant solutions to the elliptic
Boyer-Finley equation were obtained first by D. Calderbank and P.
Tod \cite{CalTod} and later, independently, in \cite{msw} where we
had also proved non-invariance of these solutions. Here, for
lifting to the solution manifold of $HCMA$, we use noninvariant
solutions to the hyperbolic version of the Boyer-Finley equation
that we have obtained in \cite{msw}  by our version of the method
of group foliation \cite{ns}. Using these solutions, we construct
explicitly metrics with ultra-hyperbolic signature that have no
Killing vectors. We hope that by an appropriate modification of
this method we shall be able to obtain noninvariant solutions of
the elliptic $CMA$ and the corresponding Ricci-flat metrics with
Euclidean signature and no Killing vectors. The other possibility
is to obtain such metrics by a suitable analytic continuation
directly from our ultra-hyperbolic metrics. A survey of results on
four-dimensional anti-self-dual metrics with the ultra-hyperbolic
signature was given by M. Dunajski in \cite{dunaj}.

In section \ref{sec-reduct}, for the sake of completeness,  we
show how a rotational symmetry reduction of $HCMA$, combined with
a point and Legendre transformations, yields the Boyer-Finley
equation.

In section \ref{sec-partner-hcma} we introduce partner symmetries
for the hyperbolic and elliptic complex Monge-Amp\`ere equations
and derive simplified equations for the case when the two partner
symmetries coincide. This is possible only for the hyperbolic
$CMA$.

In section \ref{sec-legendre} we apply a Legendre transformation
combined with a simple point transformation, similar to the one in
section \ref{sec-reduct} but with no symmetry reduction, to $HCMA$
and equations for partner symmetries. We show that, with the
choice of rotational partner symmetries, a certain linear
combination of $HCMA$ and two independent differential constraints
resulting from this choice coincides with the hyperbolic version
of the Boyer-Finley equation. Moreover, the two constraints, taken
by themselves, yield the B\"{a}cklund transformations for the
Boyer-Finley equation that we had discovered earlier \cite{mnsw}.

In section \ref{sec-lift} we use this fact for lifting
noninvariant solutions of the Boyer-Finley equation, that we had
obtained in \cite{msw}, to solutions of the $HCMA$ equation. Up to
arbitrary symmetry transformations, we give a complete list of
solutions of $HCMA$ that can be obtained by a lift from our
noninvariant solutions of the Boyer-Finley equation.

In section \ref{sec-noninv} we show that our solutions of $HCMA$
are generically noninvariant. This means that, apart from a very
special choice of arbitrary functions in these solutions, there is
no symmetry of $HCMA$ with respect to which these solutions will
be invariant. We present an explicit check of the non-invariance
for the simplest one of our solutions.

In section \ref{sec-metr-legendr} we consider four-dimensional
hyper-K\"ahler metrics with the ultra-hyperbolic signature that
are (anti-)self-dual solutions of Einstein equations provided the
metric potential satisfies $HCMA$. We introduce the tetrad of the
Newman-Penrose moving co-frame that provides an easiest and most
convenient way to calculate Riemann curvature two-form. We apply
the combination of a point and Legendre transformation, mentioned
above, to the metric and moving co-frame, so that our exact
solutions can serve as metric potentials for the transformed
metric and moving co-frame.

Finally, in section \ref{sec-new-metr} we use our solutions to
obtain explicitly new ultra-hyperbolic metrics and the
corresponding moving co-frames. Proceeding in a similar way to
\cite{mnsh}, one may check that since our solutions for the metric
potentials are noninvariant, the resulting metrics have no Killing
vectors. By utilizing the moving co-frames, we were able to
compute Riemann curvature two-forms for our solutions using the
package EXCALC (Exterior Calculus of Modern Differential Geometry)
\cite{excalc} in the computer algebra system REDUCE 3.8
\cite{reduce}.

\section{Rotational symmetry reduction of {\mathversion{bold}
$HCM\!A$} to Boyer-Finley equation}
\setcounter{equation}{0}
\label{sec-reduct}

The Boyer-Finley equation is obtained by a rotational symmetry
reduction from the elliptic $CMA$ \cite{bf}. A hyperbolic version
of the Boyer-Finley equation appears as a result of symmetry
reduction of $HCMA$ (\ref{hcma}) with respect to the group of
rotations in $(x,y)$ plane or, in the complex coordinates $z_1=x +
i y$, rotations in the complex $z_1$-plane with the generator
\[ X = y\partial_x - x\partial_y = i(\bar z_1\partial_{\bar z_1} - z_1\partial_{z_1} ) \]
and the symmetry characteristic \cite{olv} of the form $\varphi =
i(z_1u_1-\bar z_1u_{\bar 1})$. Rotationally invariant solutions
are determined by the condition $\varphi = 0$ which is satisfied
by $u = u(r,z_2,\bar z_2)$ where $r = z_1\bar z_1 = x^2 + y^2$.
For such $u$, depending only on three variables, $HCMA$ reduces to
\[ (ru_{rr} + u_r)u_{2\bar 2} - ru_{r2}u_{r\bar 2} = -1. \]
Under the change of the invariant variable $\rho = \ln{r} =
\ln{z_1} + \ln{\bar z_1}$ the reduced equation becomes
\begin{equation}
u_{\rho\rho}u_{2\bar 2} - u_{\rho 2}u_{\rho\bar 2} = - e^\rho.
\label{redeq}
\end{equation}
The Legendre transformation
\begin{equation}
u_\rho = p,\quad \rho = \phi_p,\quad u = p\phi_p - \phi,\quad z_2
= z,\quad \bar z_2 = \bar z
\label{leg}
\end{equation}
to the new unknown $\phi = \phi(p,z,\bar z)$ takes the equation
(\ref{redeq}) to the form
\begin{equation}
\phi_{z\bar z} = e^{\phi_p}\phi_{pp} \equiv (e^{\phi_p})_p
\label{bf}
\end{equation}
that is related by $F = \phi_p$ to the hyperbolic version of the
Boyer-Finley equation
\[ F_{z\bar z} = (e^F)_{pp}. \]

\section{Partner symmetries of complex\\ Monge-Amp\`ere equations}
\setcounter{equation}{0} \label{sec-partner-hcma}

  The determining equation for symmetries of $HCMA$ is
the same as for the elliptic $CMA$ (\ref{cma})
\begin{equation}
\Box(\varphi)=0,\qquad  \Box = u_{2\bar 2}D_1D_{\bar 1}+u_{1\bar
1}D_2D_{\bar 2} - u_{2\bar 1}D_1D_{\bar 2}-u_{1\bar 2}D_2D_{\bar
1}
\label{desym}
\end{equation}
where $\varphi$ denotes a symmetry characteristic and $D_i,
D_{\bar i}$ are operators of the total derivatives with respect to
$z^i, \bar z^i$ respectively. Therefore the construction of
partner symmetries, given in this section, is the same for both
elliptic and hyperbolic $CMA$ \cite{mns,mnsh}.

Define the operators
\begin{equation}
L_1 = \lambda (u_{1\bar 2}D_{\bar 1} - u_{1\bar 1}D_{\bar 2}),
\quad L_2 = \lambda (u_{2\bar 2}D_{\bar 1} - u_{2\bar 1}D_{\bar
2})
\label{L_i}
\end{equation}
where $\lambda$ is a complex constant. Then the operator $\Box$ of
the symmetry condition (\ref{desym}) can be expressed in terms of
$L_1, L_2$ as $\Box = \lambda^{-1}(D_1L_2 - D_2L_1)$. The symmetry
condition takes the form of a total divergence
\begin{equation}
D_1L_2\varphi = D_2L_1\varphi
\label{divsym}
\end{equation}
so that there locally exists a symmetry potential $\psi$ defined
by the differential equations
\begin{equation}
\psi_1 = L_1\varphi = \lambda (u_{1\bar 2}\varphi_{\bar 1} -
u_{1\bar 1}\varphi_{\bar 2}),\quad \psi_2 = L_2\varphi = \lambda
(u_{2\bar 2}\varphi_{\bar 1} - u_{2\bar 1}\varphi_{\bar 2}).
\label{psi}
\end{equation}
Because of the relation
\[ [L_1,L_2] = \lambda^2 \{(u_{1\bar 1}u_{2\bar 2} - u_{1\bar
2}u_{2\bar 1})_{\bar 1}D_{\bar 2} - (u_{1\bar 1}u_{2\bar 2} -
u_{1\bar 2}u_{2\bar 1})_{\bar 2}D_{\bar 1} \}\] the operators
$L_1$ and $L_2$ commute on solution manifolds of the elliptic and
hyperbolic $CMA$. Furthermore, we note the relation
\begin{equation}\label{rel}
 D_1L_2 - D_2L_1 = L_2D_1 - L_1D_2.
\end{equation}
Therefore, substituting $\varphi$ by its potential $\psi$ into the
symmetry condition in the divergence form (\ref{divsym}) and using
the definition (\ref{psi}) and the relation (\ref{rel}), we obtain
\[ D_1L_2\psi - D_2L_1\psi = L_2\psi_1 - L_1\psi_2 = [L_2,L_1]\psi = 0 \]
and so the potential $\psi$ of a symmetry $\varphi$ is itself a
symmetry. These two symmetries are called {\it partner
symmetries}.

Now take complex conjugate to the equations (\ref{psi}) and solve
them algebraically with respect to the $\varphi_1$ and
$\varphi_2$, using elliptic or hyperbolic $CMA$, to obtain the
inverse transformation
\begin{equation}
\varphi_1 = \mp\bar\lambda^{-1} (u_{1\bar 2}\psi_{\bar 1} -
u_{1\bar 1}\psi_{\bar 2}),\quad \varphi_2 = \mp\bar\lambda^{-1}
(u_{2\bar 2}\psi_{\bar 1} - u_{2\bar 1}\psi_{\bar 2})
\label{phi}
\end{equation}
where the minus and plus signs correspond to the elliptic and
hyperbolic $CMA$ respectively. Note that for the $HCMA$ there is a
simple possibility $\psi = \varphi$ when the equations (\ref{psi})
and (\ref{phi}) coincide and become
\begin{equation}
\varphi_1 = \lambda (u_{1\bar 2}\varphi_{\bar 1} - u_{1\bar
1}\varphi_{\bar 2}),\quad \varphi_2 = \lambda (u_{2\bar
2}\varphi_{\bar 1} - u_{2\bar 1}\varphi_{\bar 2})
\label{ps=fi}
\end{equation}
if $|\lambda| = 1$, i.e. $\lambda = e^{i\alpha}$ with a real
$\alpha$. For the elliptic $CMA$ the ansatz $\psi = \varphi$ leads
to a contradiction and no other obvious similar simplifications
exist.

We will also need the equations, complex conjugate to
(\ref{ps=fi})
\begin{equation}
\varphi_{\bar 1} = \lambda^{-1} (u_{2\bar 1}\varphi_{1} - u_{1\bar
1}\varphi_{2}),\quad \varphi_{\bar 2} = \lambda^{-1} (u_{2\bar
2}\varphi_{1} - u_{1\bar 2}\varphi_{2}).
\label{barphi}
\end{equation}
We note that any three equations for partner symmetries out of the
four ones (\ref{ps=fi}) and (\ref{barphi}) imply the fourth
equation together with $HCMA$ itself as their algebraic
consequences and, alternatively, the pair of first equations in
(\ref{ps=fi}) and (\ref{barphi}) together with $HCMA$ imply the
couple of second equations in these formulas. Thus, we have only
three independent equations. For our future needs we choose $HCMA$
together with the first equations in (\ref{ps=fi}) and
(\ref{barphi})
\begin{equation}
\varphi_1 = \lambda (u_{1\bar 2}\varphi_{\bar 1} - u_{1\bar
1}\varphi_{\bar 2}),\quad \varphi_{\bar 1} = \lambda^{-1}
(u_{2\bar 1}\varphi_{1} - u_{1\bar 1}\varphi_{2})
 \label{phi1b1}
\end{equation}
as the basic independent equations.

\section{Legendre transform of rotational partner symmetries and Boyer-Finley equation}
\setcounter{equation}{0}
\label{sec-legendre}

Next we apply to $HCMA$ and the equations (\ref{phi1b1}) the same
combination of the point transformation and Legendre
transformation, that produced the Boyer-Finley equation (\ref{bf})
in section \ref{sec-reduct} by the rotational symmetry reduction,
but now we do not perform any symmetry reduction.

The point transformation $z_1 = e^{\zeta_1}$, $\bar z_1 =
e^{\bar\zeta_1}$ yields $HCMA$ in the form
\begin{equation}
\label{zeta} u_{\zeta_1\bar\zeta_1}u_{2\bar 2} - u_{\zeta_1\bar
2\bar\zeta_12} = - e^{\zeta_1 + \bar\zeta_1}
\end{equation}
and the partner symmetries equations (\ref{phi1b1}) become
\begin{eqnarray}\label{zetpart}
 & & \varphi_{\zeta_1} = \lambda
  e^{-\bar\zeta_1}(u_{\zeta_1\bar 2}\varphi_{\bar\zeta_1} -
  u_{\zeta_1\bar\zeta_1}\varphi_{\bar\zeta_2})\nonumber
\\ & & \varphi_{\bar\zeta_1} =
\lambda^{-1}e^{-\zeta_1}(u_{2\bar\zeta_1}\varphi_{\zeta_1} -
u_{\zeta_1\bar\zeta_1}\varphi_{2}).
\end{eqnarray}

The Legendre transformation in the first pair of variables
$\zeta_1, \bar\zeta_1$
\begin{equation}
\label{legendre} \zeta_1 = \psi_q,\; \bar\zeta_1 = \psi_{\bar
q},\; u = q\psi_q + \bar q\psi_{\bar q} - \psi,\; u_{\zeta_1} =
q,\; u_{\bar\zeta_1} = \bar q ,
\end{equation}
with $z_2 = z, \bar z_2 = \bar z$, maps the unknown
$u(\zeta_1,\bar\zeta_1,z_2,\bar z_2)$ to the new unknown
$\psi(q,\bar q,z,\bar z)$ and the symmetry characteristic
transforms as $\varphi(\zeta_1,\bar\zeta_1,z_2,\bar z_2) =
\Phi(q,\bar q,z,\bar z)$. The inverse transformation is
\begin{equation}
\label{inverse}
q = u_{\zeta_1},\; \bar q = u_{\bar\zeta_1},\; \psi =
\zeta_1u_{\zeta_1} + \bar\zeta_1u_{\bar\zeta_1} - u,\; \psi_q =
\zeta_1,\; \psi_{\bar q} = \bar\zeta_1 .
\end{equation}
Under this transformation $HCMA$ (\ref{zeta}) becomes
\begin{equation}\label{leghcma}
  \psi_{q\bar q}\psi_{z\bar z} - \psi_{q\bar z}\psi_{\bar qz} =
  e^{\psi_q + \psi_{\bar q}} (\psi_{q\bar q}^2 - \psi_{qq}\psi_{\bar q\bar q})
\end{equation}
and the partner symmetries constraints (\ref{zetpart}) take the
form
\begin{eqnarray}\label{legsym}
 & &   e^{\psi_{\bar q}}(\psi_{\bar q\bar q}\Phi_q - \psi_{q\bar q}\Phi_{\bar q})
  = \lambda (\psi_{q\bar q}\Phi_z - \psi_{q\bar z}\Phi_{\bar
  q})
  \\ & &  e^{\psi_{q}}(\psi_{qq}\Phi_{\bar q} - \psi_{q\bar q}\Phi_{q})
  = \lambda^{-1} (\psi_{q\bar q}\Phi_{\bar z} - \psi_{\bar
  qz}\Phi_{q}).\nonumber
\end{eqnarray}

We use here the rotational symmetry characteristic $\varphi =
i(z_1u_1 - \bar z_1u_{\bar 1}) = i(u_{\zeta_1} - u_{\bar\zeta_1})$
with the Legendre transform $\Phi = i(q - \bar q)$ resulting from
(\ref{legendre}). This choice of $\Phi$ does not affect the
Legendre-transformed $HCMA$ (\ref{leghcma}), while the transformed
differential constraints  (\ref{legsym}), that select particular
solutions of $HCMA$ (\ref{leghcma}), become
\begin{equation}\label{legrot}
e^{\psi_{\bar q}}(\psi_{\bar q\bar q} + \psi_{q\bar q}) = \lambda
\psi_{q\bar z},\quad \lambda e^{\psi_{q}}(\psi_{qq} + \psi_{q\bar
q}) = \psi_{\bar qz} .
\end{equation}
Now, we express $\psi_{q\bar z}$ and $\psi_{\bar qz}$ from the
latter equations and substitute them into $HCMA$ (\ref{leghcma})
with the result
\begin{equation}\label{BF}
  \psi_{z\bar z} = e^{\psi_q + \psi_{\bar q}} (\psi_{qq} + 2\psi_{q\bar q} + \psi_{\bar q\bar
  q})
\end{equation}
that can be considered as a linear combination of the three
equations (\ref{leghcma}) and (\ref{legrot}). In the real
coordinates $x, y$ in the complex $q$-plane ($q=x + i y$, $\bar
q=x - i y$) the equation (\ref{BF}) becomes
\begin{equation}\label{realBF}
\psi_{z\bar z} = e^{\psi_x}\psi_{xx}
\end{equation}
which is the same Boyer-Finley equation (\ref{bf}), that we have
derived in section \ref{sec-reduct} by the rotational symmetry
reduction, but in the different variables: instead of $p$ we now
have $x=(q+\bar q)/2$. The partner symmetries constraints
(\ref{legrot}) in the real coordinates $x, y$ take the form
\begin{equation}\label{rotxy}
 \psi_{zx} + i\psi_{zy} = 2\lambda\left[e^{(\psi_x -
  i\psi_y)/2}\right]_x,\quad
  \psi_{\bar zx} - i\psi_{\bar zy} = 2\lambda^{-1}\left[e^{(\psi_x
  + i\psi_y)/2}\right]_x .
\end{equation}

 The variable $y$ does not appear explicitly in the Boyer-Finley
equation (\ref{realBF}), being just a parameter, and so it can be
regarded as a parameter of a symmetry group of this equation: a
change of $y$ will not affect the equation. If $\omega$ is the
symmetry characteristic of the Boyer-Finley equation
\begin{equation}\label{B_F}
  \tilde{\psi}_{z\bar z} = \exp{(\tilde\psi_{xx})}
\end{equation}
related to (\ref{realBF}) by the substitution $\psi =
\tilde\psi_x$, then the symmetry characteristic of (\ref{realBF})
is $i\omega_x$, where the constant factor $i$ is introduced for
convenience. The Lie equation for the symmetry group with the
parameter $y$ and symmetry characteristic $i\omega_x$ reads
\begin{equation}\label{Lie}
  \psi_y = i\omega_x .
\end{equation}
By eliminating $\psi_y$ in (\ref{rotxy}) with the aid of
(\ref{Lie}) and then integrating the resulting equations with
respect to $x$, we obtain
\begin{equation}\label{Backlund}
  \omega_z = \psi_z - 2\lambda e^{(\psi_x + \omega_x)/2},\quad
  \omega_{\bar z} = -\psi_{\bar z} + 2\lambda^{-1} e^{(\psi_x -
  \omega_x)/2} .
\end{equation}
These are B\"{a}cklund transformations for the Boyer-Finley
equation that we discovered earlier \cite{mnsw}. The differential
compatibility condition of the system (\ref{Backlund})
$(\omega_z)_{\bar z} = (\omega_{\bar z})_z$ reproduces the
Boyer-Finley equation (\ref{realBF}) and the compatibility
condition, taken in the form $(\psi_z)_{\bar z} = (\psi_{\bar
z})_z$ yields the determining equation for symmetry
characteristics of the Boyer-Finley equation (\ref{B_F})
\begin{equation}\label{omeg}
  \omega_{z\bar z} - e^{\psi_x} \omega_{xx} = 0 .
\end{equation}

Thus, without any symmetry reduction being done, the Boyer-Finley
equation arises as a linear combination of the
Legendre-transformed $HCMA$ and differential constraints
(\ref{legrot}) following from the choice of the rotational
symmetry for partner symmetries. Furthermore, the differential
constraints themselves turn out to be the B\"{a}cklund
transformations for the Boyer-Finley equation in a new disguise.

Note what happens if we reverse our procedure. Then starting with
the three-dimensional Boyer-Finley equation together with its
B\"{a}cklund transformations and considering a symmetry group
parameter $\tau$ as the fourth coordinate $y$ in the equations, we
arrive at the four-dimensional $HCMA$ equation. In this way,
partner symmetries provide a lift from three-dimensional
noninvariant solutions of the Boyer-Finley equation to
four-dimensional noninvariant solutions of $HCMA$ that govern
four-dimensional ultra-hyperbolic metrics without Killing vectors.

P. Tod in \cite{tod} used invariant solutions to both hyperbolic
Boyer-Finley equation and (\ref{omeg}) for constructing
scalar-flat K\"{a}ler metrics with ultra-hyperbolic signature that
admit a symmetry. Earlier C. LeBrun used the elliptic Boyer-Finley
equation together with the equation for its symmetries for
constructing self-dual metrics with Euclidean signature
\cite{lebrun}. Using our B\"acklund transformations, we can obtain
new noninvariant solutions of the Boyer-Finley equation, both
elliptic and hyperbolic, starting from known symmetries (solutions
to (\ref{omeg})). This approach was demonstrated in the elliptic
case in \cite{mnsw}.

\section{Lift of noninvariant solutions of the Boyer-Finley equation to {\mathversion{bold}
$HCMA$}}
 \setcounter{equation}{0}
 \label{sec-lift}

We start with noninvariant solutions to the hyperbolic version of
Boyer-Finley equation
\begin{equation}\label{b_f}
  v_{z\bar z} = (e^v)_{xx}
\end{equation}
that we had obtained earlier by the method of group foliation in
\cite{msw} (noninvariant solutions to the elliptic Boyer-Finley
equation were obtained in \cite{msw,CalTod}). Those solutions
involve a couple of holomorphic and anti-holomorphic functions of
one argument $b(z)$ and $\bar b(\bar z)$ that arise as "constants"
of integrations. In our construction, the Boyer-Finley equation
(\ref{realBF}) and its solutions depend also on the fourth
variable, the parameter $y$, and hence the integration "constants"
in the noninvariant solutions given in \cite{msw}, $b$ and $\bar
b$, also should depend on $y$:
\begin{equation}\label{sol_v}
  v(x,y,z,\bar z) = \ln{[x+b(z,y)]} + \ln{[x+\bar b(\bar z,y)]} -
2\ln{(z+\bar z)}.
\end{equation}
The Boyer-Finley equations in the forms (\ref{realBF}) and
(\ref{b_f}) are related to each other by the substitution $v =
\psi_x$ and hence solutions of (\ref{realBF}) are obtained by
integrating (\ref{sol_v}) with respect to $x$ with the "constant"
of integration $F(z,\bar z,y)$ that depends on the other three
variables:
\begin{eqnarray}\label{sol_psi}
 \psi & = & [x+b(z,y)]\ln{[x+b(z,y)]} + [x+\bar b(\bar z,y)]\ln{[x+\bar b(\bar
  z,y)]}\nonumber
  \\ & & \mbox{} - 2x[\ln{(z+\bar z)} + 1] + F(z,\bar z,y).
\end{eqnarray}
The unknown $y$-dependence in (\ref{sol_psi}) is determined by the
requirement that $\psi$ should also satisfy the
Legendre-transformed $HCMA$ (\ref{leghcma}), since we need
solutions of the latter equation.

Thus, we substitute the expression (\ref{sol_psi}) for $\psi$ in
$HCMA$ (\ref{leghcma}) and, since all the $x$-dependence is known
explicitly, it splits into several equations, corresponding to
groups of terms with a different dependence on $x$. We were able
to solve these equations and make a complete analysis of all
possible solutions.\vspace{2mm}

{\bf List of solutions:}\vspace{1mm}
\begin{eqnarray}\label{sol1}
 & & \psi = [q + b(z)]\ln{[q + b(z)]} + [\bar q + \bar b(\bar
z)]\ln{[\bar q + \bar b(\bar z)]}  \nonumber
\\ & & - (q + \bar q)[\ln{(z + \bar z)}
+ 1] + \int\!\!\int \frac{b(z) + \bar b(\bar z)}{(z + \bar
z)^2}\,d z d\bar z + r(y),
 \\  & & \psi = [q + b(z)]\ln{[q +
b(z)]} + [\bar q + \bar b(\bar z)]\ln{[\bar q + \bar b(\bar z)]}
\nonumber
\\ & & - (q + \bar q)[\ln{(z + \bar z)}
+ 1] + \int\!\!\int \frac{b(z) + \bar b(\bar z)}{(z + \bar
z)^2}\,d z d\bar z \nonumber
\\ & & \mbox{} + 2iy\ln{\left(\frac{\bar z}{z}\right)} + r(y),
\label{sol2}
\\   & & \psi = [q + b(z)]\ln{[q +
b(z)]} + [\bar q + \bar b(\bar z)]\ln{[\bar q + \bar b(\bar z)]}
\nonumber
\\ & & - (q + \bar q)[\ln{(z + \bar z)} + 1] + \int\!\!\int \frac{b(z) + \bar
b(\bar z)}{(z + \bar z)^2}\,d z d\bar z \nonumber
 \\ & & + 2i\int\ln{\left[\frac{\bar z + 2ik(y)}{z - 2ik(y)} \right]}\, d y +
r(y).
\label{sol3}
\end{eqnarray}
Here $r(y)$ and $k(y)$ are arbitrary smooth real-valued functions
of one real variable $y=i(\bar q - q)/2$ and $b(z)$ and $\bar
b(\bar z)$ are arbitrary holomorphic and anti-holomorphic
functions of one complex variable that arise when the
$y$-dependence of $b(z,y)$ and $\bar b(\bar z,y)$ is completely
determined. Solution (\ref{sol2}) is a particular simple case of
the more general solution (\ref{sol3}) when $k(y)=0$.
\begin{theorem}
Up to an arbitrary symmetry transformation of $HCMA$
(\ref{leghcma}), the list of solutions (\ref{sol1}) - (\ref{sol3})
is a complete set of solutions of $HCMA$ that can be obtained by
lifting noninvariant solutions (\ref{sol_psi}) of the Boyer-Finley
equation.
\end{theorem}

Note that, by construction, we have obtained the solutions of
$HCMA$ that satisfy only one additional differential constraint,
the Boyer-Finley equation, though in (\ref{legrot}) we have two
constraint equations produced by partner symmetries. If we require
that both constraints (\ref{legrot}) should be satisfied, then we
shall obtain a subset of solutions that are invariant with respect
to non-local symmetries of $HCMA$, though this does not mean
invariant solutions in the usual sense \cite{mns,mnsh}. Solutions
with such special property are obtained by setting $r(y)$ to be
particular linear functions, namely for solution (\ref{sol1})
\begin{equation}\label{r(y)1}
  r(y) = 2(\alpha - \pi)y + r_0
\end{equation}
and for solutions (\ref{sol2}) and (\ref{sol3})
\begin{equation}\label{r(y)23}
  r(y) = 2\alpha y + r_0
\end{equation}
where $r_0$ is an arbitrary real constant and $\lambda =
e^{i\alpha}$ is the constant coefficient in (\ref{legrot}).

\section{Non-invariance of solutions}
\setcounter{equation}{0} \label{sec-noninv}

We have found point symmetries of the Legendre-transformed $HCMA$
(\ref{leghcma}) using computer packages CRACK and LIEPDE by Thomas
Wolf \cite{wolf}, being run in the computer algebra system REDUCE
3.8. The symmetry generators are
\begin{eqnarray}\label{symgen}
  X_1 = q\partial_q + \bar q\partial_{\bar q} + (q + \bar q +
  \psi)\partial_\psi , \quad X_2 = (q - \bar q)\partial_\psi
  \\ X_{a(z)} = a(z)\partial_z - a'(z)q\partial_\psi, \quad
  X_{c(z)} = c(z)\partial_q, \quad X_{d(z)} = d(z)\partial_\psi
  \nonumber
\end{eqnarray}
together with the complex conjugate generators $\bar X_{\bar
a(\bar z)}$, $\bar X_{\bar c(\bar z)}$, and $\bar X_{\bar d(\bar
z)}$, where $a(z)$, $c(z)$, and $d(z)$ and their complex
conjugates are arbitrary holomorphic and anti-holomorphic
functions respectively. Here $\partial_q =
\partial/\partial_q$ and so on.

A solution $\psi = f(q,\bar q,z,\bar z)$ is invariant under a
one-parameter symmetry Lie group with the generator $X$ if it
satisfies the invariance condition
\begin{equation}\label{invcond}
  X(f - \psi)|_{\psi = f} = 0
\end{equation}
where, after acting by $X$ on the solution manifold, $\psi$ should
be eliminated by using the solution $\psi = f$.

In our problem, a generator of an arbitrary one-dimensional
symmetry subgroup, that should be used in the invariance condition
(\ref{invcond}), is a linear combination of the basis generators
(\ref{symgen})
\begin{equation}\label{X}
  X = C_1 X_1 + C_2 X_2 + X_{a(z)} + \bar X_{\bar a(\bar z)} + X_{c(z)}
  + \bar X_{\bar c(\bar z)}+ X_{d(z)} + \bar X_{\bar d(\bar z)}
\end{equation}
where $C_1$ and $C_2$ are arbitrary real constants and constant
coefficients of other generators are absorbed in the arbitrary
functions.

We apply the invariance condition (\ref{invcond}) to each of our
solutions (\ref{sol1})--(\ref{sol3}) and expect that this
condition will either require certain specializations of arbitrary
functions $b(z)$, $\bar b(\bar z)$, $r(y)$, and $k(y)$ or give the
result that $X=0$, that is, there is no symmetry of
(\ref{leghcma}) with respect to which the solution will be
invariant. That would mean that our solutions are generically
noninvariant.

With the generator $X$ defined by (\ref{X}), the invariance
condition for our first and simplest solution (\ref{sol1}) has the
form (primes denote derivatives)
\begin{eqnarray}
 & & \hspace*{-16pt} [C_1 + C_2 - a'(z)]q + [C_1 - C_2 - {\bar
a}'(\bar z)]\bar q + d(z) + \bar d(\bar z) + C_1\Biggl\{
b(z)\ln{[q + b(z)]} \nonumber
 \\ & & \hspace*{-16pt} \mbox{} + \bar b(\bar z)\ln{[\bar q + \bar b(\bar z)]}
 - (q + \bar q)[\ln{(z + \bar z)} - 1] + r(y) + \int\!\!\int\frac{b(z) + \bar b(\bar z)}{(z + \bar z)^2}\,d z
  d\bar z \Biggr\} \nonumber
  \\ &  & =\,  c(z)\ln{[q + b(z)]} + \bar c(\bar z)\ln{[\bar q + \bar b(\bar z)]}
  \label{invcond1}
  \\ & & \mbox{} - [c(z) + C_1q] [\ln{(z + \bar z)}
  + (i/2)r'(y)] - [\bar c(\bar z) + C_1\bar q] [\ln{(z + \bar
  z)}
  \nonumber
  \\ & & \mbox{} - (i/2)r'(y)] + a(z)\left\{ b'(z)\Bigl[\ln{\Bigl(q + b(z)\Bigr)} + 1\Bigr]
  + \int\frac{b(z) + \bar b(\bar z)}{(z + \bar z)^2}d\bar z
  \right\}\nonumber
  \\ & & \hspace*{-12pt} \mbox{} + \bar a(\bar z)\left\{ {\bar b}'(\bar z)\Bigl[\ln{\Bigl(\bar q + \bar b(\bar z)\Bigr)} + 1\Bigr]
  + \int\frac{b(z) + \bar b(\bar z)}{(z + \bar z)^2}d z
  \right\} - \frac{(q + \bar q)[a(z) + \bar a(\bar z)]}{z + \bar
  z} \nonumber
\end{eqnarray}
where $y = i(\bar q - q)/2$. Differentiating this equation twice
with respect to $z$ and $\bar z$ and splitting the resulting
equation in $q$ and $\bar q$, we arrive at the equation
\begin{equation}\label{a_eq}
  (z + \bar z)[a'(z) + {\bar a}'(\bar z)] - 2[a(z) + \bar a(\bar
  z)]= 0
\end{equation}
with the simple consequence $a''(z) + {\bar a}''(\bar z) = 0$.
After the separation of $z$ and $\bar z$, this yields $a''(z) =
2i\beta$ and ${\bar a}''(\bar z) = -2i\beta$, where $\beta$ is an
arbitrary real constant. Obtaining $a(z)$ and ${\bar a}(\bar z)$
by integration and substituting the result into (\ref{a_eq}), we
get
\begin{equation}\label{a_expr}
  a(z) = i\beta z^2 + C_3z + i\gamma,\quad \bar a(\bar z)
  = -i\beta{\bar z}^2 + C_3\bar z - i\gamma
\end{equation}
where $C_3$ and $\gamma$ are also arbitrary real constants.

Next, we differentiate the invariance condition (\ref{invcond1})
twice, first with respect to $q$, obtaining
\begin{eqnarray}
& &\hspace*{-6.5pt} \frac{a(z)b'(z) - C_1b(z) + c(z)}{q + b(z)} -
i C_1 r'(y) + (1/4)r''(y)[\bar c(\bar z) - c(z) + C_1(\bar q - q)]
\nonumber
  \\ & & \mbox{} - \frac{a(z) + \bar a(\bar z)}{z + \bar
z} + a'(z) - 2C_1 - C_2 = 0
  \label{difq}
\end{eqnarray}
and then with respect to $\bar q$ with the result
\begin{equation}\label{d_qbq}
  i r'''(y) [\bar c(\bar z) - c(z) + C_1(\bar q - q)] +
  6C_1r''(y) = 0.
\end{equation}

Differentiating (\ref{d_qbq}) with respect to $z$ or $\bar z$, we
obtain the conditions
\begin{equation}\label{cases}
  c'(z)r'''(y) = 0, \qquad {\bar c}'(\bar z)r'''(y) = 0
\end{equation}
that imply the following two cases: $r'''(y) = 0$ and $r'''(y) \ne
0$.

{\bf Case 1:}\hspace{2mm}
\begin{equation}\label{case1}
r'''(y) = 0\quad \Longrightarrow\quad r(y) = \lambda y^2 + \mu y +
\nu
\end{equation}
with real coefficients. Splitting (\ref{difq}) and its complex
conjugate in $q$, $\bar q$, we arrive at the relations
\begin{equation}\label{b_rel}
  a(z)b'(z) + c(z) - C_1b(z) = 0,\quad \bar a(\bar z){\bar b}'(\bar z)
  + \bar c(\bar z) - C_1\bar b(\bar z) = 0
\end{equation}
\begin{equation}\label{c(z)}
  (\lambda/2)[\bar c(\bar z) - c(z)] + i\beta (z + \bar z) -
  (i\mu + 2)C_1 - C_2 = 0
\end{equation}
where in the last equation we have used the expressions
(\ref{a_expr}) for $a(z)$ and $\bar a(\bar z)$, and
\begin{equation}\label{1ab}
  C_1r''(y) = 0 \quad\iff\quad C_1\lambda = 0
\end{equation}
so that either $C_1 = 0$ or $\lambda = 0$.

{\bf Case 1a:}\hspace{2mm}
\begin{equation}\label{case1a}
\lambda \ne 0,\qquad C_1 = 0.
\end{equation}
Separating $z$ and $\bar z$ in (\ref{c(z)}), we determine $c(z)$,
$\bar c(\bar z)$ and $C_2$ (using that $C_2$ is real)
\begin{equation}\label{c_barc}
  c(z) = (2i\beta/\lambda)z + c_0,\quad \bar c(\bar z) =
  -(2i\beta/\lambda)\bar z + \bar c_0 ,\quad C_2 = 0.
\end{equation}
The relations (\ref{b_rel}) become
\begin{equation}\label{b_det}
  a(z)b'(z) + c(z) = 0,\quad \bar a(\bar z){\bar b}'(\bar z)
  + \bar c(\bar z) = 0
\end{equation}
 with $a(z)$, $\bar a(\bar z)$ given by (\ref{a_expr}) and $c(z)$, $\bar
c(\bar z)$ by (\ref{c_barc}) respectively, and so they yield the
special form of $b(z)$ and $\bar b(\bar z)$. Now, differentiating
the invariance condition (\ref{invcond1}) twice with respect to
$z$ and $\bar z$ and using (\ref{b_rel}), we obtain $\beta = 0,
c_0 = 0$ and hence $c(z) = \bar c(\bar z) = 0$, so that
eliminating the trivial case when $b$ and $\bar b$ are constants,
we conclude that $a(z) = 0$ and $\bar a(\bar z) = 0$. Then the
invariance condition (\ref{invcond1}) reduces to $d(z) + \bar
d(\bar z) = 0$ and thus the symmetry generator (\ref{X}) is zero.
Therefore, in the case of quadratic $r(y)$ there is no symmetry
with respect to which our solution with the nonconstant $b$ and
$\bar b$ could be invariant.

{\bf Case 1b:}\hspace{2mm}
\begin{equation}\label{case1b}
\lambda = 0,\qquad C_1 \ne 0 \quad \Longrightarrow\quad r(y) = \mu
y + \nu.
\end{equation}
The relation (\ref{c(z)}) in this case is split in $z$ and $\bar
z$ to give
\begin{equation}\label{beta0}
  \beta = 0,\qquad \mu = 0 \quad{\rm or} \quad C_1 = 0, \qquad C_2 = -2C_1
\end{equation}
because $C_1$ and $C_2$ are real. If $C_1 = 0$, we are back to
Case 1a, so $\mu = 0$ and from (\ref{case1b}) $r(y) = \nu$ should
be constant for an invariant solution. Invariance condition
(\ref{invcond1}) with the use of the relations (\ref{b_rel}),
after splitting in $\bar q$, yields $C_1 = 0$, so we are again
back to the Case 1a.

{\bf Case 2:}\hspace{2mm}
\begin{equation}\label{case2}
r'''(y) \ne 0\quad \Longrightarrow\quad c'(z) = 0,\quad {\bar
c}'(\bar z) = 0
\end{equation}
where we have used (\ref{cases}), so $c(z) = c$ and $\bar c(\bar
z) = \bar c$ are now constants.

{\bf Case 2a:}\hspace{2mm} $C_1 \ne 0$.

  Then (\ref{d_qbq}) is easily integrated to yield
\begin{equation}\label{r(y)}
  r(y) = \frac{r_0}{8C_1^2[2C_1y + i (\bar c - c)]} + r_1y +
  r_2
\end{equation}
where $r_0,r_1$ and $r_2$ are constants of integrations.
Substituting the expression (\ref{r(y)}) into invariance condition
(\ref{invcond1}) and splitting the resulting equation in $q$ and
$\bar q$, we obtain
\begin{equation}\label{r_0}
  r_0 = 0,\quad \beta = 0,\quad C_1 = 0,\quad C_2 = 0
\end{equation}
that contradicts the assumption of the Case 2a.

{\bf Case 2b:}\hspace{2mm} $C_1 = 0$.

  Then (\ref{d_qbq}) yields $\bar c = c$ and splitting
(\ref{difq}) in $q$ and the complex conjugate to (\ref{difq}) in
$\bar q$, we obtain
\begin{equation}\label{b(z)}
a(z)b'(z) = - c,\quad \bar a(\bar z){\bar b}'(\bar z) = - c,\quad
\beta = 0,\quad C_2 = 0
\end{equation}
so that $a$ and $\bar a$ are linear functions
\begin{equation}\label{a(z)}
 a(z) = C_3z + i\gamma,\quad \bar a(\bar z)
 = C_3\bar z - i\gamma .
\end{equation}
The invariance condition (\ref{invcond1}) simplifies to
\begin{eqnarray}
& & d(z) + \bar d(\bar z) = -2c[\ln{(z + \bar z)} + 1] \nonumber
 \\ & & \mbox{} +
  a(z)\int\frac{b(z) + \bar b(\bar z)}{(z + \bar z)^2}\,d\bar z
  +  \bar a(\bar z)\int\frac{b(z) + \bar b(\bar z)}{(z + \bar
  z)^2}d z.
   \label{simple}
\end{eqnarray}
The term with the logarithm cannot be compensated by the integral
terms. Indeed, the only possibility for the integrals to produce
$\ln{(z + \bar z)}$ is when $b = kz$, $\bar b = k\bar z$ with the
real constant $k$ while $a,\bar a$ are constant (at $C_3 = 0$),
but then the integral terms cancel each other. Therefore, the
coefficient of the logarithm should vanish, so $c = 0$ and the
relations (\ref{b(z)}) yield $a = \bar a = 0$ for nonconstant
$b(z)$ and $\bar b(\bar z)$. It follows then from (\ref{simple})
that $d(z) + \bar d(\bar z) = 0$ and the symmetry generator $X$ in
(\ref{X}) vanishes. Thus, apart from the case of constant $b$ and
$\bar b$, there is no symmetry under which our solution
(\ref{sol1}) would be invariant.

For our more complicated solutions (\ref{sol2}) and (\ref{sol3})),
it is obvious that invariance conditions would be even more
difficult to satisfy and hence we can summarize our results as
follows.
\begin{theorem}
If the functions $b(z)$, $\bar b(\bar z)$ are not constants, the
formulas (\ref{sol1}) - (\ref{sol3}) yield noninvariant solutions
of $HCMA$ (\ref{leghcma}).
\end{theorem}
Note that for non-invariance of solution (\ref{sol1}) the
condition of the theorem\ \thetheorem\ is necessary and
sufficient.

As a consequence, the ultra-hyperbolic metrics governed by the
potentials $\psi$ in (\ref{sol1}) - (\ref{sol3}), constructed in
the next sections, have no symmetries (Killing vectors).

\section{Ultra-hyperbolic hyper-K\"ahler metrics,\\ Newman-Penrose co-frame
and Legendre transformation} \setcounter{equation}{0}
\label{sec-metr-legendr}

 Four-dimensional hyper-K\"ahler metrics
\begin{equation}\label{metr}
  d s^2 = u_{1\bar 1} d z^1 d\bar z^1 + u_{1\bar 2} d z^1 d\bar z^2 +
  u_{2\bar 1} d z^2 d\bar z^1 + u_{2\bar 2} d z^2 d\bar z^2
\end{equation}
satisfy Einstein field equations with either Euclidean or
ultra-hyperbolic signature, if the K\"ahler potential $u$
satisfies elliptic (\ref{cma}) or hyperbolic (\ref{hcma}) complex
Monge-Amp\`ere equation respectively \cite{pleb}. Such metrics are
Ricci-flat and have (anti-)self-dual curvature. Here we restrict
ourselves to the $HCMA$ equation (\ref{hcma}) and hence the metric
(\ref{metr}) has ultra-hyperbolic signature. This becomes obvious
if we use the tetrad of Newman-Penrose moving co-frame $\{l,\bar
l,m,\bar m\}$ \cite{gol,an} corresponding to the metric
(\ref{metr})
\begin{equation}\label{tetrad}
  l = \frac{1}{\sqrt{u_{1\bar 1}}}\big(u_{\bar 11}d z^1 + u_{\bar 12}
  d z^2 \big), \qquad m = \frac{1}{\sqrt{u_{1\bar 1}}}\, d z^2
\end{equation}
where $\bar l$ and $\bar m$ are complex conjugates to $l$ and $m$.
Indeed, if we express $u_{2\bar 2}$ from the equation (\ref{hcma})
and substitute this in the metric (\ref{metr}), then the metric
can be written in the form
\begin{equation}\label{invmetr}
  d s^2 = l \otimes \bar l - m \otimes \bar m
\end{equation}
so that the signature of the metric is ultra-hyperbolic $(++--)$.
The Newman-Penrose co-frame provides most convenient way of
calculating Riemann curvature two-forms.

Because of the discrete symmetry $1\leftrightarrow 2, \bar 1
\leftrightarrow\bar 2$ of $HCMA$ and the metric (\ref{metr}),
another possible co-frame tetrad is obtained from (\ref{tetrad})
by this discrete transformation
\begin{equation}\label{tetrad2}
  l' = \frac{1}{\sqrt{u_{2\bar 2}}}\big(u_{\bar 21}d z^1 + u_{\bar 22}
  d z^2 \big), \qquad m' = \frac{1}{\sqrt{u_{2\bar 2}}}\, d z^1
\end{equation}
and it also satisfies the relation (\ref{invmetr})
\begin{equation}\label{invmetr2}
  d s^2 = l' \otimes \bar l' - m' \otimes \bar m' .
\end{equation}

Since we have exact solutions of $HCMA$ that was subjected to a
combination of a point and Legendre transformation, in order to
use these solutions, we have to perform the same transformations
upon the metric (\ref{metr}) and the Newman-Penrose co-frame
(\ref{tetrad}).

The point transformation $z_1 = e^{\zeta_1}$, $z_2 = e^{\zeta_2}$
leaves the metric form-invariant
\begin{equation}\label{zetametr}
d s^2 = u_{\zeta_1\bar\zeta_1} d\zeta^1 d\bar\zeta^1 +
u_{\zeta_1\bar 2} d\zeta^1 d\bar z^2 +
  u_{2\bar\zeta_1} d z^2 d\bar\zeta^1 + u_{2\bar 2} d z^2 d\bar z^2.
\end{equation}
The tetrad $1$-forms become
\begin{equation}\label{transtetrad}
  l = \frac{u_{\zeta_1\bar\zeta_1} d\zeta^1 + u_{2\bar\zeta_1}
  d z^2}{\sqrt{u_{\zeta_1\bar\zeta_1}}}\,, \qquad
  m = \frac{e^{(\zeta_1 +
  \bar\zeta_1)/2}d z^2}{\sqrt{u_{\zeta_1\bar\zeta_1}}}
\end{equation}
together with their complex conjugates $\bar l, \bar m$, where we
have skipped the exponential factors $e^{(\zeta_1 -
\bar\zeta_1)/2}$ and $e^{(\bar\zeta_1 - \zeta_1)/2}$ in $l$ and
$\bar l$ since they cancel each other in the formula
(\ref{invmetr}) for the metric.

Next we perform the Legendre transformation (\ref{legendre}) of
the metric and moving co-frame. The metric becomes
\begin{eqnarray}
 d s^2 & = & \frac{-1}{\Delta_-}\,\Bigl\{\psi_{qq}(\psi_{\bar qq}d q + \psi_{\bar qz}d z)^2
 + \psi_{\bar q\bar q}(\psi_{q\bar q}d\bar q + \psi_{q\bar z}d\bar
 z)^2
 \nonumber
 \\ & + & \Delta_+( \psi_{q\bar q}d qd\bar q + \psi_{q\bar z}d qd\bar z + \psi_{\bar qz}d\bar q d z
 + \psi_{z\bar z}d zd\bar z )
 \label{legmetr}
\\ & + & 2\psi_{q\bar q}(\psi_{q\bar z}\psi_{\bar qz} - \psi_{q\bar q}\psi_{z\bar z})
d z d\bar z \Bigr\} \nonumber
\end{eqnarray}
where $\Delta_- = \psi_{qq}\psi_{\bar q\bar q} - \psi_{q\bar
q}^2$, $\Delta_+ = \psi_{qq}\psi_{\bar q\bar q} + \psi_{q\bar
q}^2$ and $\psi(q,\bar q,z,\bar z)$ has to satisfy
(\ref{leghcma}), the Legendre transform of $HCMA$. The Legendre
transform of the moving co-frame is
\begin{eqnarray}\label{legtetrad}
 & & l = \frac{\psi_{q\bar q}(\psi_{qq}d q + \psi_{q\bar q}d\bar q + \psi_{q\bar z}d\bar z)
  + \psi_{qq}\psi_{\bar qz}d z}{\sqrt{-\psi_{q\bar
  q}\Delta_-}}\nonumber
  \\ & & m = e^{(\psi_q + \psi_{\bar q})/2}\sqrt{\frac{-\Delta_-}{\psi_{q\bar
  q}}}\,d z
\end{eqnarray}
together with their complex conjugates. It is easy to check that
these $d s^2$, $l$, and $m$ together with $\bar l$ and $\bar m$
still satisfy the relation (\ref{invmetr}). The Legendre transform
of the co-frame (\ref{tetrad2}) is
\begin{eqnarray}
 & l' = \Bigl\{(\psi_{q\bar q}\psi_{\bar q\bar z} - \psi_{\bar q\bar q}\psi_{q\bar z})
 (\psi_{qq}d q + \psi_{q\bar q}d\bar q + \psi_{q\bar z}d\bar z) \nonumber
 \\ & \mbox{} + [\psi_{\bar qz}(\psi_{qq}\psi_{\bar q\bar z} - \psi_{q\bar q}\psi_{q\bar z}) - \psi_{z\bar z}\Delta_-]d z \Bigr\}
 \times
\nonumber
 \\ & \left\{ \Delta_-\,[\psi_{qz}(\psi_{q\bar z}\psi_{\bar q\bar q} - \psi_{q\bar q}\psi_{\bar q\bar z})
 + \psi_{\bar qz}(\psi_{qq}\psi_{\bar q\bar z} - \psi_{q\bar q}\psi_{q\bar z}) - \psi_{z\bar z}\Delta_-]
 \right\}^{-1/2}
 \nonumber
 \\ & m' = e^{\psi_q}\sqrt{\Delta_-}\,(\psi_{qq}d q + \psi_{q\bar q}d\bar q + \psi_{qz}d z + \psi_{q\bar z}d\bar
 z)\times
 \label{legtetrad2}
 \\ & \left\{\psi_{qz}(\psi_{q\bar z}\psi_{\bar q\bar q} - \psi_{q\bar q}\psi_{\bar q\bar z})
 + \psi_{\bar qz}(\psi_{qq}\psi_{\bar q\bar z} - \psi_{q\bar q}\psi_{q\bar z}) - \psi_{z\bar
 z}\Delta_- \right\}^{-1/2} \nonumber
\end{eqnarray}
and their complex conjugates. These $l'$, $m'$, $\bar{l'}$,
$\bar{m'}$, and $d s^2$ satisfy the relation (\ref{invmetr2}).

\section{New ultra-hyperbolic metrics and moving co-frames}
\setcounter{equation}{0} \label{sec-new-metr}

To obtain new ultra-hyperbolic Ricci-flat metrics without Killing
vectors together with moving co-frames, we use for $\psi$ in the
formulas (\ref{legmetr}) and (\ref{legtetrad}) our noninvariant
solutions of $HCMA$ (\ref{leghcma}) from the list (\ref{sol1}) -
(\ref{sol3}).

For the first solution (\ref{sol1}) from this list, the metric
takes the form
\begin{eqnarray}
 & &  d s^2 = \frac{- 4}{(z + \bar z)^2[(q + \bar q + b + \bar
b)r''(y) +  4]} \left\{ (\sqrt{A}d q - \sqrt{D}d z)^2 \right.
\label{metr1}
\\ & & \left. \mbox{} + (\sqrt{\bar A}d\bar q  - \sqrt{\bar D}d\bar z)^2  + B\left[d
qd\bar q
 \pm\sqrt{D/A}(d qd\bar z + d\bar qd z) \right] + E
  d zd\bar z \right\}
 \nonumber
\end{eqnarray}
where the plus or minus sign corresponds to $r''(y)>0$ or
$r''(y)<0$ respectively and the metric coefficients are defined by
the formulas
\[ D = \frac{1}{4}\left(\bar q + \bar b\right)\left[\left(q + b\right)r''(y) + 4\right],
\quad A = \frac{1}{16}\,(z + \bar z)^2 (r''(y))^2 D \] together
with their complex conjugates and
\begin{eqnarray*}
 & & B = -\frac{1}{32}\left(z + \bar z\right)^2r''(y)\left[(q +
b) \left(\bar q + \bar b\right)(r''(y))^2 \right. \\
 & & \left. \mbox{} + 2 \left(q + \bar q + b + \bar b\right)r''(y) + 8
 \right]
\end{eqnarray*}
\[ E = \frac{1}{4}\left[ (q + b)^2 + (\bar q + \bar b)^2
 \right]r''(y) + q + \bar q + b + \bar b . \]
 From now on, $b=b(z)$ and $\bar b = \bar b(\bar z)$ are arbitrary
 holomorphic and anti-holomorphic functions of one complex argument and $r(y)$ is an
 arbitrary real-valued function of one real variable $y = i(\bar q - q)/2$.

 The calculation of the affine connection
one-forms and the curvature two-forms is greatly facilitated by
the use of the Newman-Penrose moving co-frame \cite{gol,an}. For
generic solutions we shall use the first, simpler co-frame $l,
 \bar l, m, \bar m$ defined by (\ref{legtetrad}).

 The co-frame forms for the first solution are
\begin{eqnarray}
 & & \hspace*{-52.5pt} l = \frac{(z + \bar z)(q + b)(r''(y))^2(d\bar q - d q) + 4(q +
b)r''(y)(d\bar z - d z) - (z + \bar z)r''(y)d q - d z}{4(z + \bar
z)\{r''(y)[(q + \bar q + b + \bar b)r''(y) + 4]\}^{1/2}} \nonumber
\\ & & m = \left\{\frac{(q + \bar q + b + \bar b)r''(y)
+ 4}{r''(y)}\right\}^{1/2}\,\frac{d z}{z + \bar z} \label{tetr1}
\end{eqnarray}
and the complex conjugates $\bar l, \bar m$.

For the second solution (\ref{sol2}), the metric becomes
\begin{eqnarray}
 & & d s^2 = 4\frac{(\sqrt{A}d q - \sqrt{D}d z)^2 + (\sqrt{\bar
A}d\bar q - \sqrt{\bar D}d\bar z)^2}{z^2\bar z^2(z + \bar z)^2[4 -
(q + \bar q + b + \bar b)r''(y)]} \nonumber
\\ & & \mbox{} + 4B \frac{[z(z + \bar z)r''(y)d q + 4\bar zd z]
[\bar z(z + \bar z)r''(y)d\bar q + 4zd\bar z]}{z^3\bar z^3(z +
\bar z)^4(r''(y))^2[4 - (q + \bar q + b + \bar b)r''(y)]}
\nonumber
\\ & & \mbox{} + \frac{[4 - (q + \bar q + b + \bar b)r''(y)]}{(z + \bar
z)^2r''(y)}\, d zd\bar z
 \label{metr2}
\end{eqnarray}
where the metric coefficients are defined by the formulas
\[ D = \frac{1}{4}\,\bar z^4(\bar q + \bar b)[(q + b)r''(y) - 4],\quad
A = \left(\frac{z}{4\bar z}\right)^2(z + \bar z)^2 (r''(y))^2 D\]
together with their complex conjugates and
\begin{eqnarray*}
 & & \hspace*{-16pt} B = -\frac{z^2\bar z^2}{32}\,(z + \bar z)^2r''(y)[(q + b) (\bar
q + \bar b)(r''(y))^2  - 2(q + \bar q + b + \bar b)r''(y) + 8].
\end{eqnarray*}
The co-frame tetrad becomes
\begin{eqnarray}
  & & l = [4z\bar z(z + \bar z)]^{-1} \left\{\frac{\bar q + \bar
  b}{(q + b)r''(y)[(q + \bar q + b + \bar b)r''(y) - 4] }\right\}^{1/2}\times
  \nonumber
  \\ & & \biggl\{ z\bar z(z + \bar z)r''(y)\bigl[(q + b)r''(y)(d q - d\bar q) - 4d q
  \bigr]
  \nonumber
  \\ & & \mbox{} - 4(q + b)r''(y)(z^2d\bar z - \bar z^2d z) - 16
  \bar z^2d z  \biggr\}
  \nonumber
  \\ & & m = \left\{\frac{(q + \bar q + b + \bar b)r''(y) -
  4}{r''(y)}\right\}^{1/2} \, \frac{d z}{z + \bar z}
    \label{tetr2}
\end{eqnarray}
together with $\bar l$, $\bar m$.

For the third solution (\ref{sol3}), generalizing (\ref{sol2}), we
use a shorthand notation
\begin{eqnarray}
  & & V = ( z + \bar z)k'(y) - \frac{1}{4}\,(z - 2i k(y))(\bar z
  + 2i k(y))r''(y)  \nonumber
  \\ & & W = (q + b)V + (z - 2i k(y))(\bar z  + 2i k(y))
  \label{not}
\end{eqnarray}
and $\bar W$ is the complex conjugate to $W$. Here $k = k(y)$ and
$r(y)$ are arbitrary smooth real-valued functions of a real
variable $y = i(\bar q - q)/2$ that appear in the third solution
(\ref{sol3}). The metric has the form
\begin{eqnarray}
 & & \hspace*{-20pt} d s^2 = \nonumber
 \\ & & \mbox{} \hspace*{-20pt} - \Bigl\{(z + \bar z)^2(z - 2i k)^2(\bar z + 2i k)^2[(q
+ \bar q + b + \bar b)V + (z - 2i k)(\bar z + 2i k)]\Bigr\}^{-1}
\nonumber
  \\ & & \times \biggl\{ (\bar q + \bar b)W\Bigl[(\bar z + 2i k)^2d z - (z + \bar z)Vd q\Bigr]^2
  \nonumber
 \\ & & \mbox{} + (q + b)\bar W\Bigl[(z - 2i k)^2d\bar z - (z + \bar z)Vd\bar q\Bigr]^2
  + \Bigl[W\bar W + (q + b)(\bar q + \bar b)V^2\Bigr]
  \nonumber
  \\ & & \times\Bigl[ -(z + \bar z)^2Vd qd\bar q + (z + \bar z)\Bigl((z - 2i k)^2d qd\bar z
  + (\bar z + 2i k)^2d \bar q d z\Bigr)
  \nonumber
  \\ & &  + (q + \bar q + b + \bar b)(z - 2i k)(\bar z + 2i k)d zd\bar z \Bigr] \biggr\}
  \nonumber
  \\ & & \mbox{} + \frac{2(q + b)(\bar q + \bar b)V}{(z + \bar z)^2(z - 2i k)(\bar z + 2i k)}
  \,d zd\bar z
  \label{metr3}
\end{eqnarray}
and the co-frame $1$-forms are
\begin{eqnarray}
 & & \hspace*{-13pt} l = \frac{ (z + \bar z)V\Bigl[ \bar Wd\bar q - (\bar q + \bar
b)Vd q \Bigr] + (\bar q + \bar b)(\bar z + 2i k)^2Vd z - (z - 2i
k)^2\bar Wd\bar z } { (z + \bar z)(\bar z + 2i k)^2\{V[(q + \bar q
+ b + \bar b)V + (z - 2i k)(\bar z + 2i k)]\}^{1/2} }\nonumber
\\ & & m = \left\{\frac{(q + \bar q + b + \bar b)V + (z
- 2i k)(\bar z + 2i k)}{V}\right\}^{1/2}\, \frac{d z}{z + \bar z}
  \label{tetr3}
\end{eqnarray}
and the complex conjugates $\bar l$, $\bar m$.

By utilizing the moving co-frames, we were able to compute Riemann
curvature two-forms for our solutions using the package EXCALC
(Exterior Calculus of Modern Differential Geometry) \cite{excalc}
in the computer algebra system REDUCE 3.8 \cite{reduce}.

The special solutions (\ref{sol1}) and (\ref{sol2}) meeting the
restrictions (\ref{r(y)1}) and (\ref{r(y)23}) respectively, that
satisfy both constraints (\ref{legrot}), are simple enough to
enable us to present explicitly metrics, moving co-frames, and
Riemann curvature tensors. For the first special solution
(\ref{sol1}) with the restriction (\ref{r(y)1}), the metric reads
\begin{equation}\label{metric1}
 d s^2 = (z + \bar z)^{-2}\{ (z + \bar z)(d qd\bar z
  + d\bar qd z) - \Bigl[\Bigl(q + b(z)\Bigr)d\bar z
  + \Bigl(\bar q + \bar b(\bar z)\Bigr)d z\Bigr]
  (d z + d\bar z) \}.
\end{equation}
For the second special solution (\ref{sol2}) with the restriction
(\ref{r(y)23}), the metric is
\begin{eqnarray}
 & & d s^2 = [z\bar z(z + \bar z)]^{-2}\Bigl\{ \Bigl[\Bigl(\bar q + \bar b(\bar z)\Bigr){\bar z}^2d z
 + \Bigl(q + b(z)\Bigr)z^2d\bar z\Bigr]\Bigl({\bar z}^2d z + z^2d\bar z\Bigr) \nonumber
 \\ & & \mbox{} + z\bar z(z + \bar z)\Bigl(z^2d qd\bar z + {\bar z}^2d\bar q d z\Bigr)
 \Bigr\}.
   \label{metric2}
\end{eqnarray}
Both of these metrics are Ricci-flat and have only one
non-vanishing component of the  Riemann curvature tensor
\begin{equation}\label{curv1}
 R_{3434} = 2^{-1}(z + \bar z)^{-3}\Bigl\{ 2\Bigl[
b'(z) + {\bar b}'(\bar z) \Bigr] - (z + \bar z)\Bigl[ b''(z) +
{\bar b}''(\bar z) \Bigr] \Bigr\}
\end{equation}
for the first special solution and
\begin{eqnarray}
& & R_{3434} = \Bigl(2z^2{\bar z}^2(z + \bar
z)^3\Bigr)^{-1}\Bigl\{ (z + \bar z)\Bigl[z^4 b''(z) + {\bar
z}^4{\bar b}''(\bar z) \Bigr] \nonumber
  \\ & & \mbox{} + 2z^3(z + 2\bar z)b'(z) +
2{\bar z }^3(\bar z + 2z){\bar b}'(\bar z) \Bigr\}
  \label{curv2}
\end{eqnarray}
for the second special solution. For the Riemann tensor
$R^a_{\phantom{a}bcd}$ there are two non-vanishing components
\[ R^2_{\phantom{2}434} = - R^1_{\phantom{2}334} = 2(z + \bar z)R_{3434} \]
for the first special solution and
\[ R^2_{\phantom{2}434} = \frac{2z}{\bar z}\,R_{3434},\quad R^1_{\phantom{2}334} = \frac{2\bar z}{z}\,R_{3434} \]
for the second special solution.

For the two special solutions the first moving co-frame
(\ref{legtetrad}) becomes singular because of the vanishing
$r''(y)$ (and hence $\psi_{q\bar q}$) in the denominators. There
is no such difficulty for the more general solution (\ref{sol3})
with the restriction (\ref{r(y)23}). Therefore, for the two
special solutions we have to use the second co-frame $l',
 \bar{l'}, m', \bar{m'}$ defined by (\ref{legtetrad2}).
For the first special solution it becomes
\begin{eqnarray}
 & & l' = (z + \bar z)^{-3/2}[-(b' + \bar b')]^{-1/2}\{(z +
 \bar z)(dq - \bar b'dz) - (q + b)(dz + d\bar
 z)\}
 \nonumber
 \\ & & m' = (z + \bar z)^{-3/2}[-(b' + \bar b')]^{-1/2}[(q + b)(dz + d\bar
 z) - (z + \bar z)(dq + b'dz)]\nonumber
 \\
\label{frame2sol1}
\end{eqnarray}
together with complex conjugates. For the second special solution
the second moving co-frame reads
\begin{eqnarray}
 & & \hspace*{-10.3pt} l' = \Bigl\{(q + b)z\bigl(\bar z^2dz + z^2d\bar z\bigr) + \bar z(z + \bar z)
 \bigl\{ \bigr[ 2\bar z(\bar q + \bar b) - \bar z^2\bar b' \bigl]dz + z^2dq \bigr\}
 \Bigr\}\times
 \nonumber
\\ & & z^{-1/2}[\bar z(z + \bar z)]^{-3/2}\{ 2[z(q + b) + \bar z(\bar q + \bar b)] - (z^2b' + \bar z^2\bar b') \}
 \nonumber
\\ & & \hspace*{-10.3pt} m' = \sqrt{\bar z}[z(z + \bar z)]^{-3/2}
\bigl\{(q + b)[(2z + \bar z)\bar zdz - z^2d\bar z] - z\bar z(z +
\bar z)(dq + b'dz)\bigr\} \nonumber
\\ & & \times \bigl\{ 2[z(q + b) + \bar z(\bar q + \bar b)] - (z^2b' + \bar z^2\bar b') \bigl\}
  \label{frame2sol2}
\end{eqnarray}
and their complex conjugates.

Using these co-frames with the package EXCALC we were able to
compute Riemann curvature two-forms for both solutions. For the
first special solution they read
\begin{eqnarray}
 & & R^2_{\phantom{2}2} = \frac{2}{b' + \bar b'}\left(\frac{b'' +
\bar b''}{b' + \bar b'} - \frac{2}{z + \bar z} \right) \times
\nonumber
\\ & & \bigl[ {\rm o(2)}\wedge {\rm o(3)} - {\rm o(1)}\wedge {\rm o(4)} -
{\rm o(3)}\wedge {\rm o(4)} - {\rm o(1)}\wedge {\rm o(2)} \bigr]
 \nonumber
\\ & & R^3_{\phantom{2}1} = R^3_{\phantom{2}3} =
R^2_{\phantom{2}4} = - R^1_{\phantom{2}1} = - R^4_{\phantom{2}2} =
- R^1_{\phantom{2}3} = - R^4_{\phantom{2}4} = R^2_{\phantom{2}2}
\nonumber
\\ & & R^1_{\phantom{2}2} = R^1_{\phantom{2}4} = R^2_{\phantom{2}1} =
R^2_{\phantom{2}3}= R^3_{\phantom{2}2} = R^3_{\phantom{2}4} =
R^4_{\phantom{2}1} = R^4_{\phantom{2}3} = 0.
 \label{riem1}
\end{eqnarray}
From now on we use the notation ${\rm o(1)} = l'$, ${\rm o(2)} =
\bar l'$, ${\rm o(3)} = m'$, and ${\rm o(4)} = \bar m'$ for the
co-frame tetrads.

For the second special solution Riemann curvature two-forms are
\begin{eqnarray}
 & & R^1_{\phantom{1}1} = (z\bar z)^{-2}(z + \bar z)^{-1}\Bigl\{
z^2b' + \bar z^2\bar b' - 2[z(q + b) + \bar z(\bar q + \bar b)]
\Bigr\}^{-2}\times \nonumber
\\ & & \Bigl[ (z + \bar z)(z^4b'' + \bar z^4\bar b'')
+ 2z^3(z + 2\bar z)b' + 2\bar z^3(2z + \bar z)\bar b'\Bigr]\times
 \label{riem2}
 \\ & & \Bigl\{ z^4\,{\rm o(2)}\wedge {\rm o(3)} - \bar z^4\,{\rm o(1)}\wedge {\rm o(4)}
 - (z\bar z)^2 [{\rm o(1)}\wedge {\rm o(2)} + {\rm o(3)}\wedge {\rm o(4)}] \Bigr\}
 \nonumber
 \\ & & R^2_{\phantom{1}2} = R^3_{\phantom{1}3} = R^4_{\phantom{1}4} =
 - R^1_{\phantom{1}1},\; R^1_{\phantom{1}3} =
 R^4_{\phantom{1}2} = \frac{z^2}{\bar
 z^2}\,R^1_{\phantom{1}1},\; R^2_{\phantom{1}4} = -
 R^3_{\phantom{1}1} = \frac{\bar
 z^2}{z^2}\,R^1_{\phantom{1}1}\nonumber
 \\ & & R^1_{\phantom{2}2} = R^1_{\phantom{2}4} = R^2_{\phantom{2}1} =
R^2_{\phantom{2}3}= R^3_{\phantom{2}2} = R^3_{\phantom{2}4} =
R^4_{\phantom{2}1} = R^4_{\phantom{2}3} = 0.
  \nonumber
\end{eqnarray}

\section{Conclusions}

Our goal is to obtain noninvariant solutions of four-dimensional
heavenly equations because they will yield new gravitational
metrics with no Killing vectors. In particular, this property
characterizes the famous gravitational instanton $K3$ where the
metric potential should be a noninvariant solution of the elliptic
complex Monge-Amp\`ere equation. In this paper we have developed a
suitable approach for solving similar problem for an easier case
of the hyperbolic complex Monge-Amp\`ere equation. This approach
is based on the use of partner symmetries for lifting noninvariant
solutions of three-dimensional equations, that can be obtained
from $HCMA$ by the symmetry reduction, to non-invariant solutions
of the original four-dimensional equation.

A symmetry reduction of a partial differential equation reduces by
one the number of independent variables in the original equation,
so that the reduced equation is easier to solve. Its solutions are
solutions of the original PDE that are invariant under the
symmetry that was used in the reduction. Even if we found
noninvariant solutions of the reduced equation, it would only mean
that no further symmetry reduction was being made and they would
still be invariant solutions of the original equation. On an
example of the hyperbolic complex Monge-Amp\`ere equation, we have
shown that partner symmetries, when they exist, provide a
possibility for a procedure reverse to the symmetry reduction: a
lift of noninvariant solutions of the reduced equation to
noninvariant solutions of the original equation of higher
dimensions. We have developed such a procedure for $HCMA$ and
obtained new noninvariant solutions of this equation. Using these
solutions as metric potentials, we have obtained new gravitational
metrics with the ultra-hyperbolic signature that have no Killing
vectors. The calculation of the affine connection one-forms and
the curvature two-forms is greatly facilitated by the use of the
Newman-Penrose moving co-frame which we have calculated for all
our solutions.

We are now in the process of developing a modified lifting
procedure to apply it to the elliptic complex Monge-Amp\`ere
equation. Using new noninvariant solutions of this equation as
metric potentials, we shall obtain new gravitational metrics with
the Euclidean signature and with no Killing vectors. We hope to
obtain in such a way at least some pieces of the Kummer surface
$K3$.

\section*{Acknowledgements}

The research of MBS is partly supported by the research grant from
Bogazici University Scientific Research Fund, research project No.
07B301.


\begin{thebibliography}{99}
 \bibitem{pleb} Pleba\~nski J F 1975 {\it J.
 Math. Phys.} {\bf 16} 2395--402
 \bibitem{ahs} Atiyah M F, Hitchin N J and Singer I M 1978
{\it Proc. Roy. Soc. A} {\bf 362} 425--61
 \bibitem{mnsh}
Malykh A A, Nutku Y and  Sheftel M B 2004 {\it J. Phys. A: Math.
Gen.} {\bf 37} 7527--45 (Preprint math-ph/030503)
 \bibitem{mns} Malykh A A, Nutku Y and Sheftel M B 2003
{\it J. Phys. A:  Math. Gen.} {\bf 36} 10023--37
 \bibitem{mnsgr} Malykh A A, Nutku Y and Sheftel M B 2003
{\it Class. Quantum Grav.} {\bf 20} L263--66
 \bibitem{bf} Boyer C P and Finley III J D 1982
{\it J. Math. Phys.} {\bf 23} 1126--30
 \bibitem{CalTod}
Calderbank D M J and Tod P 2001 {\it Differ. Geom. Appl.} {\bf 14}
199--208 {\it J. Phys. A: Math. Gen.} {\bf 34} 137--56
 \bibitem{msw} Martina L, Sheftel M B and Winternitz P 2001
{\it J. Phys. A: Math. Gen.} {\bf 34}, 9243--63
 \bibitem{ns} Nutku Y and Sheftel M B 2001
\bibitem{dunaj}
Dunajski M and West S 2006 Preprint math.DG/0610280
 \bibitem{mnsw} Malykh A A, Nutku Y, Sheftel M B and Winternitz P 1998
{\it Physics of Atomic Nuclei (Yadernaya Fizika)} {\bf 61},
1986--89
 \bibitem{excalc} Schr\"ufer E 2003 {\it EXCALC: A differential geometry
 package} in: Hearn A C {\it REDUCE, User's and Contributed Packages Manual, Version 3.8},
 Ch. {\bf 39} 333--343
 \bibitem{reduce} Hearn A C 2003 {\it REDUCE, User's and Contributed Packages Manual, Version 3.8}
 \bibitem{olv} Olver P 1986 {\it Applications of Lie Groups to Differential Equations}
(New York: Springer-Verlag)
\bibitem{tod}
Tod K P 2001 in: {\it Further Advances in Twistor Theory}, Vol.
III, Chapman \& Hall/CRC, 61--63
\bibitem{lebrun}
LeBrun C 1991 {\it J. Diff. Geom.} {\bf 34} 223-53
\bibitem{wolf} Wolf T 1985 {\it J. Comp. Phys.} {\bf 60} 437--446
 \bibitem{gol} Goldblatt E 1994 {\it Gen. Rel. and Grav.} {\bf 26}
979
 \bibitem{an} Aliev A N and Nutku Y 1999 {\it Class. Quantum Gravity}
{\bf 16} 189
 \end{thebibliography}
\end{document}